\documentclass[12pt]{article}
\usepackage{graphicx}
\usepackage[cp1251]{inputenc}
\usepackage{rotating}
\begin{document}
 \noindent {\footnotesize\it Astronomy Reports, 2017, Vol. 61, No 10, pp. 883--890}
 \newcommand{\dif}{\textrm{d}}

 \noindent
 \begin{tabular}{llllllllllllllllllllllllllllllllllllllllllllll}
 & & & & & & & & & & & & & & & & & & & & & & & & & & & & & & & & & & & & & \\\hline\hline
 \end{tabular}

 \vskip 0.5cm
  \centerline{\bf\large Search for Close Stellar Encounters with the Solar System}
  \centerline{\bf\large from Data on Nearby Dwarfs}
 \bigskip
 \bigskip
  \centerline
 {
 V.V. Bobylev
 }
 \bigskip
 \centerline{\small \it
 Central (Pulkovo) Astronomical Observatory, Russian Academy of Sciences,}
 \centerline{\small \it Pulkovskoe shosse 65, St. Petersburg, 196140 Russia}
 \bigskip
 \bigskip
 \bigskip

 {
{\bf Abstract}---Trigonometric parallaxes measured with
ground-based telescopes of the RECONS consortium as part of the
CTIOPI program are used to search for stars that have either had
an encounter with the solar system in the past or will have such
an encounter in the future, at distances of less than a few
parsecs. These are mainly low-mass dwarfs and subdwarfs of types
M, L, and T currently at distances of less than 30 pc from the
Sun. Six stars for which encounters with the solar orbit at
distances of less than 1 pc are possible have been identified for
the first time. For example, the minimum distance for the star
**SOZ~3A will $0.72\pm0.11$ pc at an epoch of $103\pm44$ thousand
years in the future.
  }

\medskip DOI: 10.1134/S106377291710002X

 \section{INTRODUCTION}
The Sun is surrounded by the cometary Oort cloud. It is believed
that its radius is approximately $1\times105$ AU (0.5 pc). At such
distances, the gravitational bonds between these comets and the
Sun are weak, and their orbits can be subject to slight
perturbations due to various external factors. These factors
include the random passage of a field star, the influence of giant
molecular clouds, and the influence of the overall Galactic
gravitational field [1--3].

Passages of Galactic field stars near the Oort cloud or even
penetration into this cloud can induce the formation of cometary
showers moving in the vicinity of the giant planets [4]. The
possible bombardment of the Moon and Earth with such cometary
bodies is not ruled out [5].

Numerous studies have been dedicated to searches for close
encounters of stars with the solar orbit. About 200 HIPPARCOS
stars [6] that either underwent a close encounter with the solar
system at a distance of less than 5 pc in the past or will do so
in the future, at epochs from $-10$ to $+10$ million years, are
currently known. According to the estimates of Garcia–S\'anchez et
al. [7], the frequency of stellar encounters at distances of less
than 1 pc is roughly 12 stellar encounters per one million years.

Several specific stars have a high probability of penetrating into
the vicinity of the Oort cloud. As was first shown by
Garcia–S\'anchez et al. [8], and later confirmed by other data [7,
9, 10], the star GJ 710 (a K7V dwarf) may have a very close
encounter with the solar orbit. For example, Bobylev [9] used data
from the revised HIPPARCOS catalog to determine this star's
minimum distance to be $d_m=0.31\pm0.17$ pc at the epoch
$t_m=1447\pm60$ thousand years in the future. Finally, a
completely new estimate was made by Berski and Dybczy\'nski [11]
using parallaxes and proper motions measured by GAIA [12],
yielding $d_m=0.065\pm0.030$ pc and $t_m=1350\pm50$ thousand
years. Thus, the star GJ~710 currently holds the record for the
closest encounter.

The estimates $d_m=0.25^{+0.11}_{-0.07}$ pc and
$t_m=-70^{+0.15}_{-0.10}$ thousand years were obtained for the low
mass binary system WISE J072003.20$-$084651.2 (M9.5+T5) in [13].
The total mass of this system is only about $0.15M_\odot$. Two
more such stars are also known: HIP 85605 ($d_m\sim0.1$ pc,
$t_m\sim330$ thousand years) [14] and HIP 63721 ($d_m\sim0.2$ pc,
$t_m\sim150$ thousand years) [14, 15]. However, the trigonometric
parallaxes for these last two stars are currently quite
unreliable. Dybzhi\'nski [9] carried out numerical simulations of
the evolution of cometary orbits based on the example of the
penetration of GJ~710, with a mass of about $0.6M_\odot$, into the
Oort cloud. These simulations showed that Galactic tides can give
rise to a greater flux of comets in the inner part of the solar
system than a star with these parameters. The conclusion that a
star such as GJ~710 has a small destructive influence on the Oort
cloud was also drawn in [16]. However, the later simulations of
[11] with new parameters for a closer flyby of this star showed
the possibility of an appreciable flux corresponding to tens of
comets per year over three to four million years.

The aim of our study was to search for stars that could undergo
close encounters with the Sun, based on a sample of nearby stars
whose trigonometric parallaxes were measured using ground-based
telescopes. Radial-velocity measurements obtained in the RAVE
project [17] have recently become available for a large number of
weak stars.

 \section{DATA}
We used results obtained in the framework of the international
Research Consortium On Nearby Stars (RECONS,
http://www.astro.gsu.edu/RECONS/index.htm). Observation aimed at
deriving stellar trigonometric parallaxes and proper motions were
carried out in the Chilean Andes starting in 1999 as part of the
Cerro Tololo Interamerican Observatory Parallax Investigation
(CTIOPI), using two telescopes with mirror diameters 0.9 and
1.5~m. The mean accuracy of the trigonometric parallaxes is 3
milliarcsecond (mas).

The stars GJ 3379 (G 99$-$049) and GJ 3323 (LHS 1723) were
identified in [18] as close-encounter candidates, based on a list
of 100 nearby stars of the RECONS program. The amount of data
available for analyzes of this sort has now appreciably grown.

The latest results of trigonometric parallax measurements
conducted in the CTIOPI program are presented in [19--26]. The
total list of stars with measured parallaxes and proper motions
compiled based on these publications contains more than 500
objects, primarily M, L, and T dwarfs.

The RAdial Velocity Experiment (RAVE) [17] is dedicated to massive
determinations of radial velocities for weak stars. Observations
in the southern hemisphere on the 1.2~m Schmidt telescope of the
Anglo--Australian Observatory began in 2003. Since then, five
editions of this catalog (DR1--DR5) have been published. The mean
uncertainty in the radial velocities is about 3 km/s.

We took the radial velocities used in our current study mainly
from the RAVE DR4 catalog [27], which is accessible via the SIMBAD
electronic database. As a result, based on the 500 stars of the
initial CTIOPI list, we created a working database containing the
measured trigonometric parallax, proper motion components, and
radial velocity for each of 175 stars.

We especially dedicated this separate study to a search for close
encounters with the Sun by stars whose trigonometric parallaxes
were determined using specific ground telescopes. The reliability
of these parallaxes enables a confident analysis of only a small
circumsolar volume with a radius of no more than 25--30~pc. This
makes it possible to apply simpler analysis methods, such as
linear or epicyclic fitting.

 \section{METHOD}
 \subsection{Epicyclic Orbit}
Based on the epicyclic approximation [28], we constructed orbits
of the objects studied in coordinates rotating about the Galactic
center:
 \begin{equation}
 \renewcommand{\arraystretch}{2.2}
 \begin{array}{lll}
 \displaystyle
 x(t)=x_0+{u_0\over\kappa}\sin(\kappa t)+{v_0\over 2B}(1-\cos(\kappa t)),\\\displaystyle
 y(t)=y_0+{2\Omega_0 u_0\over\kappa^2}(1-\cos(\kappa t)) 
         +2A\biggl(x_0+{v_0\over 2B}\biggr)t
          -{\Omega_0 v_0 \over B\kappa}\sin(\kappa t),\\\displaystyle
 z(t)=z_0\cos(\nu t)+{w_0\over \nu}\sin(\nu t),
 \end{array}
 \label{epiciclic}
 \end{equation}
where $t$ is the time in millions of years (using the relation 1
pc/million years = 0.978 km/s); $A$ and $B$ are the Oort
constants; $\kappa=\sqrt{-4\Omega_0 B}$ is the epicyclic
frequency; $\Omega_0$ the angular speed of Galactic rotation of
the Local Standard of Rest (LSR), $\Omega_0=A-B$; and
$\nu=\sqrt{4\pi G\rho_0}$ the frequency of vertical oscillations,
where $G$ is the gravitational constant and $\rho_0$ the stellar
density in the solar neighborhood. Equation (1) is written for
heliocentric coordinates, where $x$ is directed away from the Sun
toward the Galactic center, $y$ points in the direction of
Galactic rotation, and $z$ points in the direction of the Galactic
north pole. The space velocities of stars $(u,v,w)$ are directed
along the axes $(x,y,z).$

The parameters $(x_0,y_0,z_0)$ and $(u_0,v_0,w_0)$ in the system
of equations (1) denote the initial positions and velocities of
the stars. The velocities $(u,v,w)$ were corrected for the
peculiar motion of the Sun relative to the LSR, using the
components $(U,V,W)_\odot=(11.1,12.2,7.3)$ km/s [29]. We took the
local density of matter to be $\rho_0=0.1~M_\odot/$pc$^3$, in
accordance with the estimate of [30], which gives $\nu=74$
km/s/kpc. We used the values of the Oort constants $A=15.5\pm0.3$
km/s/kpc and $B=-12.2\pm0.7$ km/s/kpc, found from masers with
measured trigonometric parallaxes, which corresponds to
$\Omega_0=27.7$ km/s/kpc and $\kappa=41$ km/s/kpc. The height of
the Sun above the Galactic plane was taken to be $z_0=16\pm2$~pc
[32]. We neglected the gravitational interaction between the stars
and the Sun.

 \subsection{Statistical Modeling}
In accordance with the Monte Carlo statistical modeling method, we
computed a set of orbits for each star ascribing random errors to
the input data. We computed the approach parameter for the orbits
of the stars and Sun, $d(t)=\sqrt{\Delta x^2(t)+\Delta
y^2(t)+\Delta z^2(t)}$, for each star. The epoch of minimum
approach is characterized by the two numbers $d_m$ and $t_m.$ We
assumed that the errors in the stellar parameters were normally
distributed with standard deviation $\sigma.$

In practice, this worked as follows. We found a set of model
orbits based on Eqs. (1). Random uncertainties in the equatorial
coordinates, proper motion components, parallax, and radial
velocity were added for each star. We then computed the mean
values of $d_m, t_m,$ and their standard deviations.

 \begin{table}[p]                               
 \caption[]{\small\baselineskip=1.0ex\protect
 Data for stars with approach parameters $d_m<2.3$ pc
 }
 \begin{center}
 \begin{tabular}{|r|l|r|r|r|r|r|r|r|}\hline
 \label{tab-1}
  No. &   Star   & $\alpha_{J2000}$ & $\mu_\alpha\cos\delta,\qquad\quad$ & $\pi,$    &  $V_r,$ &  $d_m,$ & $t_m,$  \\
    &            & $\delta_{J2000}$ & $\mu_\delta,$ mas/yr   &   mas     &  km/s   &    pc   &  Gyr \\\hline
  1 & L 87-10    & $ 16.028958$     & $ -323.11\pm0.71$      & $  85.40$ & $  -867$ & $   0.27$ & $   14$ \\
    &            & $-65.374250$     & $ -171.08\pm0.37$      & $\pm1.50$ & $ \pm17$ & $\pm0.01$ & $ \pm1$ \\\hline
  2 & LHS 3583   & $311.654500$     & $  540.65\pm0.71$      & $  94.72$ & $   970$ & $   0.42$ & $  -11$ \\
    &            & $-81.720472$     & $ -540.65\pm0.71$      & $\pm2.38$ & $ \pm7 $ & $\pm0.02$ & $ \pm1$ \\\hline
  3 & GJ 4274    & $335.779042$     & $  304.11\pm0.19$      & $ 137.58$ & $   308$ & $   0.62$ & $  -23$ \\
    &            & $-17.606944$     & $ -703.52\pm0.28$      & $\pm0.50$ & $\pm116$ & $\pm0.08$ & $\pm10$ \\\hline
  4 & ** SOZ 3A  & $214.204500$     & $   91.27\pm3.90$      & $ 111.15$ & $   -87$ & $   0.72$ & $  103$ \\
    &            & $ 13.807306$     & $  132.52\pm1.34$      & $\pm4.99$ & $ \pm33$ & $\pm0.11$ & $\pm44$ \\\hline
  5 & LHS 1351   & $ 32.825417$     & $ -682.47\pm1.16$      & $  71.53$ & $   953$ & $   0.74$ & $  -15$ \\
    &            & $-63.228055$     & $ -344.74\pm0.59$      & $\pm1.64$ & $ \pm11$ & $\pm0.04$ & $\pm 1$ \\\hline
  6 & LEHPM 4771 & $337.539416$     & $  -62.20\pm0.04$      & $  63.70$ & $   940$ & $   0.93$ & $  -17$ \\
    &            & $ 53.748750$     & $ -740.69\pm0.50$      & $\pm1.11$ & $ \pm10$ & $\pm0.03$ & $\pm 1$ \\\hline
  7 & LHS 500    & $313.904666$     & $ 1415.40\pm0.47$      & $  82.79$ & $   912$ & $   1.12$ & $  -13$ \\
    &            & $ 14.065277$     & $ -468.10\pm0.20$      & $\pm1.24$ & $ \pm21$ & $\pm0.06$ & $\pm 1$ \\\hline
  8 & G 99-049   & $ 90.014666$     & $  305.24\pm0.30$      & $ 193.60$ & $  30.2$ & $   1.25$ & $ -161$ \\
    &            & $  2.706555$     & $  -38.02\pm0.04$      & $\pm1.85$ & $\pm0.5$ & $\pm0.06$ & $\pm 6$ \\\hline
  9 & GJ 1157    & $185.755958$     & $ -742.72\pm0.45$      & $  62.42$ & $  -778$ & $   1.28$ & $   20$ \\
    &            & $ 46.619000$     & $ -346.34\pm0.21$      & $\pm0.63$ & $ \pm46$ & $\pm0.12$ & $\pm 1$ \\\hline
 10 & GJ 729     & $282.455708$     & $  637.86\pm0.38$      & $ 339.59$ & $ -10.5$ & $   1.95$ & $  157$ \\
    &            & $-23.836222$     & $ -192.58\pm0.12$      & $\pm1.63$ & $\pm0.1$ & $\pm0.02$ & $\pm 4$ \\\hline
 11 & GJ 1061    & $ 53.998708$     & $  733.93\pm0.24$      & $ 268.66$ & $  -20$  & $   2.18$ & $  122$ \\
    &            & $-44.512583$     & $ -368.69\pm0.33$      & $\pm0.59$ & $ \pm5$  & $\pm0.14$ & $\pm21$ \\\hline
 12 & CN Leo     & $164.120250$     & $-3808.09\pm0.30$      & $ 413.13$ & $  19.3$ & $   2.28$ & $  -14$ \\
    &            & $  7.014666$     & $-2692.61\pm0.42$      & $\pm1.27$ & $\pm0.2$ & $\pm0.02$ & $\pm 1$ \\\hline
 \end{tabular}
 \end{center}
 \end{table}
 \begin{figure}[t]
{\begin{center}
 \includegraphics[width=0.9\textwidth]{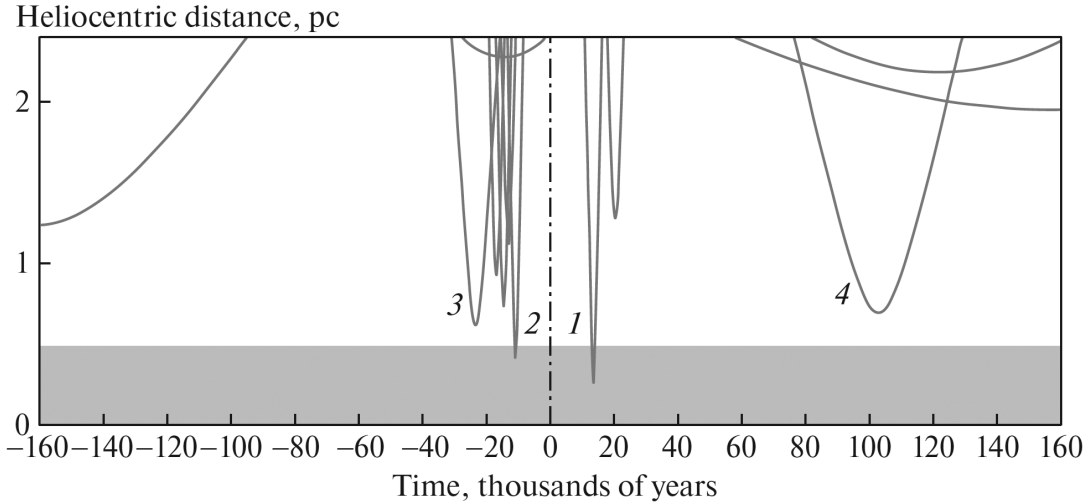}
 \caption{\small
Trajectories of stars relative to the Sun. The vertical line marks
the current epoch, grey shading indicates the boundary of the Oort
cloud, and the trajectories of four stars are numbered in
accordance with their ordinal numbers in the tables.
  } \label{f-1}
\end{center}}
\end{figure}

\section*{RESULTS AND DISCUSSION}
For each of the 175 stars in our list, we constructed its orbit
relative to the solar orbit in the interval from $–1$ million
years to $+1$ million years. It turned out that many stars with
huge radial velocities fell into the sample, which are most likely
due to erroneous measurements in the RAVE4 catalog. Therefore, we
restricted our consideration to 12 stars with approach parameters
$d_m<2.3$~pc, whose characteristics are given in Table 1. The
figure shows the approach trajectories of these stars with the
solar orbit.

The columns of Table 1 give (1) an ordinal number for the star,
(2) an identification number, (3) the equatorial coordinates
$\alpha$ and $\delta$, (4) the components of the proper motion
$\mu_\alpha\cos\delta,$ $\mu_\delta$ and their uncertainties, (5)
the trigonometric parallax $\pi$ and its uncertainty, (6) the
radial velocity $V_r$ and its uncertainty, and (7)--(8) estimates
of the approach parameters $d_m$ and $t_m.$

Table 2 gives the input coordinates and velocities of the selected
stars. Many stars from this list have very high space velocities
(exceeding 600~km/s). This would seem to indicate that they should
be classified as hypervelocity stars, capable of escaping the
gravitation of the Galaxy. The escape velocity at the Sun’s
distance from the Galactic center depends slightly on the model
used for the gravitational potential, and is about 550 km/s (see,
e.g., [33]). Due to the high speeds of these stars, their flybys
past the solar system are brief.

It is interesting that the RAVE4 catalog also gives the
signal-to-noise ratio (SNR) and flags $c_1$--$c_{20}$, describing
the morphology of the spectrum. According to these
characteristics, all the stars in our list with radial velocities
$|V_r|>300$ km/s (Table 1) have very low SNRs (less than five,
while the SNRs for ``good'' spectra should be an order of
magnitude higher), and the spectra of all of these stars have gaps
in their continua ($c_{1,2,3}$=``c'') or are peculiar
($c_{1,2,3}$=``p''). On this basis, we concluded that the
radial-velocity measurements of these stars were of poor quality.

The RAVE radial velocities for the stars considered were usually
obtained from one ``poor'' spectrum. However, the RAVE catalogs
also contain stars for which several spectra of various quality
were taken at various epochs. For example, four radial velocity
measurements are available for the star TYC 4888-146-1, derived
using four good spectra ($c_{1,2,3}$=``n'', normal spectrum). All
four values are close to $V_r=-15$ km/s. However, the value
$V_r=1897$ km/s is also presented in the RAVE5 catalog [34], found
from a spectrum with gaps in its continuum ($c_{1,2,3}$=``c'').
All this strengthens our impression that stars with very high
radial velocities in the RAVE catalogs are likely to be
problematic.

Moreover, radial velocities obtained from spectra other than those
from the RAVE program are available for the stars GJ 4274 and GJ
1157. For GJ 4274, $V_r=-2.1\pm1.1$ km/s [35], implying
$d_m\sim6$~pc. For GJ 1157, $V_r=42\pm1.1$ km/s [36], which yields
$d_m>4$~pc, appreciably different from the values in Table 1.

The next five stars in Table 1 --- **SOZ 3A, G 99--049, GJ 729, GJ
1061, and CN Leo --- have more or less reliable radial velocities.
As we can see from Table 3, their radial velocities were taken
from sources other than the RAVE catalogs.

Table 3 gives the physical characteristics of the stars. The
masses of the M dwarfs were estimated using the data of [40]. The
spectral classification of L~87--10 is known with large
uncertainty (it is not included in the standard set of data in the
SIMBAD electronic database). We estimated its spectral type to be
M4 based on its position on a color--absolute magnitude diagram
[20,26] and the color index V--K$_s$=0.868. It's spectral type is
given as M5 in [41].

Three stars were considered earlier in [18]: GJ 3379 (G 99--049),
GJ 3323 (LHS 1723), and SDSS J1416$+$1348 (**SOZ~3A). In our
current study, we adopted more reliable values for the
trigonometric parallaxes and radial velocities of GJ 3379
(G~99--049) and GJ 3323 (LHS 1723), compared to those used in
[18]. However, the derived approach parameters for these two stars
virtually coincide with those obtained earlier.

\begin{table}[t]\caption[]{\small
  Initial velocities and coordinates of selected stars }
 \begin{center}
 \label{tab-UVW}
 \small
\begin{tabular}{|r|c|r|}\hline
  No. &  $(U,V,W)\pm(e_U,e_V,e_W),$~km/s & $(x,y,z)\pm(e_x,e_y,e_z),$~pc \qquad\quad\\\hline
  1 & $( -256, ~466, ~685)\pm(  5,   9,  13)$ & $( -3.72, -6.23, -9.19)\pm(0.07, 0.11, 0.16)$ \\
  2 & $( ~519, -643, -509)\pm(  4,   5,   4)$ & $(  5.99, -6.80, -5.42)\pm(0.15, 0.17, 0.14)$ \\
  3 & $(  136,  ~92, -262)\pm( 42,  36,  79)$ & $(  3.20,  2.74, -5.92)\pm(0.01, 0.01, 0.02)$ \\
  4 & $(  -36, ~~~5,  -79)\pm( 13,   1,  30)$ & $(  3.64,  0.21,  8.23)\pm(0.16, 0.01, 0.37)$ \\
  5 & $( ~236, -543, -748)\pm(  3,   7,   9)$ & $(  2.77, -8.23, -10.96)\pm(0.06, 0.19, 0.25)$ \\
  6 & $( ~526, -267, -722)\pm(  6,   3,   8)$ & $(  8.75, -3.64, -12.51)\pm(0.15, 0.06, 0.22)$ \\
  7 & $(  587, ~402, -578)\pm( 14,  10,  11)$ & $(  8.35,  5.68,  -6.72)\pm(0.13, 0.08, 0.10)$ \\
  8 & $(-26.2,-16.8,  0.7)\pm(0.5, 0.2, 0.1)$ & $( -4.62, -2.11,  -0.91)\pm(0.04, 0.02, 0.01)$ \\
  9 & $( -393,  629, -246)\pm( 21,  39,  13)$ & $(  7.20,-13.62,  4.41)\pm(0.07, 0.11, 0.16)$ \\
 10 & $(-12.0, -1.0, -7.2)\pm(0.1, 0.0, 0.1)$ & $(  2.84,  0.57, -0.53)\pm(0.01, 0.00, 0.00)$ \\
 11 & $(    3,    0,   25)\pm(  1,   3,   4)$ & $( -0.70, -2.13, -2.97)\pm(0.01, 0.01, 0.01)$ \\
 12 & $(-27.9,-47.6,-13.7)\pm(0.1, 0.1, 0.2)$ & $( -0.59, -1.21,  2.01)\pm(0.01, 0.01, 0.01)$ \\
   \hline
\end{tabular}
\end{center}
\end{table}
 \begin{table}[t]                           
 \caption[]{\small\baselineskip=1.0ex\protect
 Additional characteristics of the stars
 }
 \begin{center}
 \begin{tabular}{|r|l|c|c|c|c|c|c|c|}\hline
 \label{tab-masses}
  No. &  Star  & Spectral & Mass,     & Source of & Source of   \\
    &          & type     & $M_\odot$ & $V_r$     & $\mu, \pi$ \\\hline
  1 & L 87--10   & $\sim$M4   & $~0.2~$   & RAVE4    & [30] \\
  2 & LHS 3583   & M2.5       & $~0.5~$   & RAVE4    & [23] \\
  3 & GJ 4274    & M4.5Ve     & $~0.2~$   & RAVE4    & [29] \\
  4 & ** SOZ 3A  & L7+T7.5    & $<0.08$   & [43]     & [29] \\
  5 & LHS 1351   & M2.5       & $~0.5~$   & RAVE4    & [23] \\
  6 & LEHPM 4771 & M4.5       & $~0.2~$   & RAVE4    & [30] \\
  7 & LHS 500    & M5         & $~0.17$   & RAVE4    & [25] \\
  8 & G 99-049   & M3.5Ve     & $~0.3~$   & [28]     & [28] \\
  9 & GJ 1157    & M4         & $~0.2~$   & RAVE4    & [23] \\
 10 & GJ 729     & M3.5Ve     & $~0.3~$   & [45]     & [28] \\
 11 & GJ 1061    & M5.5V      & $~0.15$   & [40]     & [29] \\
 12 & CN Leo     & M5.0Ve     & $~0.17$   & [44]     & [29] \\\hline
 \end{tabular}
 \end{center}
 \end{table}

The situation for **SOZ~3A is different. Previously, only a
photometric distance estimate was available ($d=8.0\pm1.6$~pc),
and other values for the component proper motions and a
substantially different radial velocity ($V_r=-42\pm5$ km/s) were
used. The approach parameters $d_m=1.24\pm0.65$~pc and
$t_m=186\pm44$ thousand years were obtained for this star in [18];
these differ substantially from the values presented in Table 1.
**SOZ~3A has a known companion, ULAS J141623.94$+$134836.3, which
is separated from the primary of the system by about 75 AU. The
companion was detected from the similarity of its proper motions
to those of the primary [42]. At present, the system is only a
suspected binary. Therefore, long spectral observations of these
stars aimed at determining their orbital characteristics and the
systemic radial velocity would be helpful.

The stars L~87--10 and LHS~3583 have a high probability $P$ of
penetrating into the vicinity of the cometary Oort cloud. We
constructed 10 000 model orbits for these stars. Since the nominal
random uncertainties in the input data are small, the cloud of
model orbits for each star occupies a very compact region in the
$d-t$ diagram. L~87--10 always falls in the vicinity of the Oort
cloud, $d_m<0.485$~pc, and has the probability $P=10 000/10 000=
1.0$ (the ratio of the number of orbits that fall in the vicinity
of the Oort cloud to the total number of model orbits). Applying
the same approach for the star LHS 3583 yields $P=0.999.$ We
assumed their masses to be approximately $0.2M_\odot$ and
$0.5M_\odot$, respectively; i.e., lower than the mass of GJ~710.
The values of $d_m$ derived for these stars do not differ strongly
from the value for GJ~710, $d_m=0.31$~pc.

On the other hand, the masses of L~87--10 and LHS~3583 differ by a
factor of two. It is interesting to compare their possible
gravitational influence on objects in the Oort cloud. To do this,
we can use the approach of [43] to estimate the radius of the
sphere of influence of the passing star Ra (the gravitational
attraction of the star dominates inside this sphere):
  \begin{equation}
 R_a= \frac{d_m}{1+\sqrt{M_\odot/M_\star}},
  \end{equation}
where $M_\odot$ is the mass of the Sun and $M_\star$ the mass of
the star. We then calculated the distance from the Sun to the
boundary, $D_0=d_m-R_a,$ beginning from which the gravitational
influence of the star on the comet dominates. As a result, we
found $D_0=0.19$ pc for L~87--10 ($M_\star=0.2 M_\odot,$
$d_m=0.27$~pc) and $D_0=0.25$~pc for LHS~3583 ($M_\star=0.2
M_\odot,$ $d_m=0.27$~pc). The boundary D0 lies inside the Oort
cloud for both stars. Moreover, this result indicates that, in
spite of the difference in mass, the gravitational influence of
L~87--10 can extend to a closer vicinity of the Sun than the
influence of LHS 3583.

Guided by the results of numerical simulations of variations of
the parameters of cometary orbits after a close flyby of a star
such as GJ 710 obtained in [9, 16], we can conclude that the
passage of the stars L 87--10 and LHS 3583 near the solar system
may not have significant visible consequences (in the sense that
it is difficult to distinguish a flux of comets created by the
action of Galactic tides on a flux initiated by the flyby of one
of these stars). The times for their close approaches with the Sun
are shorter than for GJ 710. The difference in the flyby times is
clearly visible, for example, from a comparison of the figure
presented here and [10, Fig. 2].

Our list also contains stars with more reliable (in a systematic
sense) radial velocities. Of these, the closest approach is
predicted for the low-mass system **SOZ~3A, for which an encounter
with the solar orbit at a minimum distance of $0.72\pm0.11$~pc is
expected at an epoch $103\pm44$ thousand years in the future. The
probability of penetrating into the region of the Oort cloud was
estimated from an analysis of the 10 000 model orbits; due to the
fairly large random uncertainties in the input data, this
probability is nonzero, $P = 115/10 000 = 0.115.$

\section*{CONCLUSION}
We have carried out a search for stars that have approached or
will approach the solar system to distances of less than 2 pc. We
compiled an initial list containing kinematic data for 175 stars.
The proper motions and trigonometric parallaxes of these stars
were taken from a series of publications by the RECONS consortium
via the CTIOPI program. The radial velocities of all of these
stars were taken from various literature sources. A substantial
number of the stars have radial velocities from the RAVE4 catalog,
making it possible for many stars to analyze their space
velocities for the first time.

All these stars are located within 30 pc of the Sun,and are
low-mass dwarfs of spectral types M, L, and T. Most of these stars
have large proper motions. However, there are essentially no stars
from the HIPPARCOS and Tycho-2 catalogs,making important the
high-accuracy trigonometric parallaxes and proper motions obtained
through the CTIOPI program.

We traced the position of each star relative to the solar orbit
over a time interval from $-1$ to $+1$ million years. We have
identified for the first time six stars that may approach the
solar system to distances of less than 1~pc.

Two stars have high probabilities of penetrating into the region
of the cometary Oort cloud. The first of these, L~87--10, is
predicted to approach to a minimum distance of $d_m =
0.27\pm0.01$~pc at epoch $t_m = 14\pm1$~thousand years. For the
second, LHS~3583, $d_m = 0.42\pm0.02$~pc and $t_m =
-11\pm1$~thousand years. However, the radial speeds of these stars
exceed 500 km/s, since they were obtained from spectra with very
low quality, which could contain appreciable errors.

Our list also includes stars with more reliable radial velocities.
One example is the low-mass system **SOZ~3A (the primary SDSS
J1416$+$1348 and the secondary ULAS J141623.94$+$134836.3), for
which an encounter with the solar orbit at a minimum distance of
$0.72\pm0.11$~pc at the epoch $103\pm44$ thousand years is
possible in the future.

 \subsubsection*{ACKNOWLEDGEMENTS}
We thank the referee for useful comments that helped improve this
paper. This work was supported by the Basic Research Program P--7
of the Presidium of the Russian Academy of Sciences, subprogram
``Transitional and Explosive Processes in Astrophysics.'' Much
useful information was obtained through the SIMBAD database.

 \bigskip\subsubsection*{REFERENCES}

 {\small
\quad~1. P. A. Dybczy\'nski, Astron. Astrophys. 396, 283 (2002).

 2. P. A. Dybczy\'nski, Astron. Astrophys. 441, 783 (2005).

 3. C. A. Martinez-Barbosa, L. J\'ylkov\'a, S. Portegies Zwart, and A. G. A. Brown,
    Mon. Not. R. Astron. Soc. 464, 2290 (2017).

 4. J. G. Hills, Astron. J. 86, 1730 (1981).

 5. J. T. Wickramasinghe and W. M. Napier, Mon. Not. R. Astron. Soc. 387, 153 (2008).

 6. The HIPPARCOS and Tycho Catalogues, ESA SP--1200 (ESA, Noordwijk, The Netherlands, 1997).

 7. J. Garcia-S\'anchez, P. R. Weissman, R. A. Preston, D. L. Jones, J.-F. Lestrade,
    D.W. Latham, R. P. Stefanik, and J.M. Paredes, Astron. Astrophys. 379, 634 (2001).

 8. J. Garcia-S\'anchez, R. A. Preston, D. L. Jones, P. R. Weissman, J.-F. Jean-Francois,
    D. W. Latham, and R. P. Stefanik, Astron. J. 117, 1042 (1999).

 9. P. A. Dybczy\'nski, Astron. Astrophys. 449, 1233 (2006).

 10. V. V. Bobylev, Astron. Lett. 36, 220 (2010).

 11. F. Berski and P. A. Dybczy\'nski, Astron. Astrophys. 595, L10 (2016).

 12. A. G. A. Brown, A. Vallenari, T. Prusti, J. de Bruijne, et al., Astron. Astrophys.
     595, 2 (2016).

 13. E. E. Mamajek, S. A. Barenfeld, V. D. Ivanov, A. Y. Kniazev, P. V\"ais\"anen,
     Y. Beletsky, and H. M. J. Boffin, Astrophys. J. 800, L17 (2015).

 14. C. A. L. Bailer-Jones, Astron. Astrophys. 575, 35 (2015).

 15. P. A. Dybczy\'nski and F. Berski, Mon. Not. R. Astron. Soc. 449, 2459 (2015).

 16. J. J. Jim\'enez-Torres, B. Pichardo, G. Lake, and H. Throop,
     Mon. Not. R. Astron. Soc. 418, 1272 (2011).

 17. M. Steinmetz, T. Zwitter, A. Seibert, F. G.Watson, et al., Astron. J. 132, 1645 (2006).

 18. V. V. Bobylev, Astron. Lett. 36, 816 (2010).

 19. A. R. Riedel, J. P. Subasavage, C. T. Finch, W.- C. Jao, et al., Astron. J.
    140, 897 (2010).

 20. A. R. Riedel, C. T. Finch, T. J. Henry, J. P. Subasavage, et al., Astron. J.
    147, 85 (2014).

 21. W.-C. Jao, T. J. Henry, J. P. Subasavage, J. G. Winters, A. R. Riedel, and
     P. A. Ianna,Astron. J. 141, 117 (2011).

 22. J. C. Lurie, T. J. Henry, W.-C. Jao, S. N. Quinn, et al., Astron. J. 148, 91 (2014).

 23. S. B. Dieterich, T. J. Henry, W.-C. Jao, J. G. Winters, A. D. Hosey, A. R. Riedel,
     and J. P. Subasavage, Astron. J. 147, 94 (2014).

 24. C. L. Davison, R. J. White, T. J. Henry, A. R. Riedel, et al., Astron. J. 149, 106 (2015).

 25. A. J. Weinberger, A. P. Boss, S. A. Keiser, G. Anglada-Escud\'e, I. B. Thompson,
 and G. Burley, Astron. J. 152, 24 (2016).

 26. J. G. Winters, R. A. Sevrinsky, W.-C. Jao, T. J.Henry, A. R. Riedel,
 J. P. Subasavage, J. C. Lurie, and P. A. Ianna, Astron. J. 153, 14 (2017).

 27. G. Kordopatis, G. Gilmore, M. Steinmetz, C. Boeche, et al., Astron. J. 146, 134 (2013).

 28. B. Lindblad, Arkiv Math. Astron. Fys. A 20 (17) (1927).

 29. R. Sch\"onrich, J. Binney, and W. Dehnen, Mon. Not. R. Astron. Soc. 403, 1829 (2010).

 30. J. Holmberg and C. Flinn,Mon. Not. R. Astron. Soc. 352, 440 (2004).

 31. V. V. Bobylev, A. T. Bajkova, A. S. Stepanishchev, Astron. Lett. 34, 515 (2008).

 32. V. V. Bobylev and A. T. Bajkova, Astron. Lett. 42, 1 (2016).

 33. A. T. Bajkova and V. V. Bobylev, Astron. Lett. 42, 567 (2016).

 34. A. Kunder, G. Kordopatis, M. Steinmetz, T. Zwitter, et al., Astron. J. 153, 75 (2017).

 35. J. E. Gizis, I. N. Reid and S. L. Hawley, Astron. J. 123, 3356 (2002).

 36. A. W. Rodgers and O. Eggen, Publ. Astron. Soc. Pacif. 86, 742 (1974).

 37. K. N. Abazajian, J. K. Adelman-McCarthy, M. A. Ag\"ueros, S. S. Allam, et al.,
     Astrophys. J. Suppl. 182, 543 (2009).

 38. D. L. Nidever, G. W. Marcy, R. P. Butler, D. A. Fisher, and S. S. Vogt,
     Astrophys. J. Suppl. 141, 503 (2002).

 39. R. C. Terrien, S. Mahadevan, C. F. Bender, R. Deshpande, and P. Robertson,
     Astrophys. J. 802, L10 (2015).

 40. G. F. Benedict, T. J. Henry, O. G. Franz, B. E. McArthur, et al.,
     Astron. J. 152, 141 (2016).

 41. S. Lepine and E. Gaidos, Astron. J. 142, 138 (2011).

 42. R.-D. Scholz, Astron. Astrophys. 510, L8 (2010).

 43. A. A. M\"ull\"ari and V. V. Orlov, Earth, Moon, Planets 72, 19 (1996).

 }
 \end{document}